\documentstyle[aps,preprint,prb,epsf]{revtex}

\begin{document}

\title{Statistical Physics in Meteorology}

\author{  M. Ausloos }

\address{ SUPRATECS\footnote{SUPRATECS = Services Universitaires Pour la
Recherche et les Applications Technologiques de mat\'eriaux Electroc\'eramiques,
Composites et Supraconducteurs} and GRASP\footnote{GRASP = Group for Research in
Applied Statistical Physics}, Institute of Physics, B5, \\University of Li$\grave
e$ge, B-4000 Li$\grave e$ge, Belgium}

\date{\today}


\draft

\maketitle

\abstract{Various aspects of modern statistical physics and meteorology can be
tied together. The historical importance of the University of Wroclaw in the
field of meteorology is first pointed out. Next, some basic difference about time
and space scales between meteorology and climatology is outlined. The nature and
role of clouds both from a geometric and thermal point of view are recalled.
Recent studies of  scaling laws for atmospheric variables are mentioned, like
studies on cirrus ice content, brightness temperature, liquid water path
fluctuations, cloud base height fluctuations, ....  Technical time series
analysis approaches based on modern statistical physics considerations are
outlined.}

\vskip 1cm

\section{ INTRODUCTION and FOREWORD}

This contribution to the 18th Max Born Symposium Proceedings, cannot be seen as
an extensive review of the connection between meteorology and various aspects of
modern statistical physics. Space and time (and weather) limit its content. Much
of what is found here can rather be considered to result from a biased view point
or  limited understanding of a frustrated new researcher unsatisfied by the
present status of the field.  Yet only to be found is a set of basic
considerations and reflections expecting to give lines for various
investigations,   in the spirit of modern statistical physics ideas.

The author came into this subject starting from previous work in econophysics,
when he observed that some "weather derivatives" were in use, and some sort of
game initiated by the Frankfurt Deutsche B\"{o}rse\cite{xelsius} in order to
attract customers which could predict the temperature in various cities within a
certain lapse of time, and win some prize thereafter. This subject was similar to
predicting the S\&P500 or other financial index values at a certain future time.
Whence various techniques which were used in econophysics, like the detrended
fluctuation analysis, the multifractals, the moving average crossing techniques,
etc. could be attempted from scratch.

Beside the weather (temperature) derivatives other effects are of interest. Much
is said and written about e.g. the ozone layer and the Kyoto ''agreement''. The
El Ni$\tilde{n}$o system is a great challenge to scientists. Since there is some
data available under the form of time series, like  the Southern Oscillation
Index, it is of interest to look for trends, coherent structures, periods,
correlations in noise, etc. in order to bring some knowledge, if possible basic
parameters, to this meteorological field and expect to import some modern
statistical physics ideas into such climatological phenomena. It appeared that
other data are  also available, like those obtained under various experiments,
put into force by various agencies, like   the Atlantic Stratocumulus Transition
Experiment (ASTEX) for ocean surfaces or those of the Atmospheric Radiation
Measurement Program\cite{web,arm} (ARM), among others.

However it appeared that the  data is sometimes of rather limited value because
of the lack of precision, or are biased because the raw data is already
transformed through models, and arbitrarily averaged (''filtered'') whence even
sometimes lacking the meaning it should contain. Therefore a great challenge
comes through in order to sort out the wheat from the chaff in order to develop
meaningful studies. I will mention most of the work to which I have contributed,
being aware that I am failing to acknowledge many more  important reports than
those, - for what I truly apologize. There are very interesting lecture notes on
the web for basic modules on meteorological training courses, e.g. one available
through ECMWF website\cite{ECMWF}.

In Sect.2, I will  briefly comment on the history of meteorology. The notion of
clouds, in Sect. 3, allows for bringing up  the geometrical notion of fractals
for meteorology work, thus scaling laws, and modern data analysis techniques. 
Simple technical and useful approaches, based on standard statistical physics
techniques and ideas, in particular based on the scaling hypothesis for phase
transitions and percolation theory features will be found in Sect. 4.

\section{ HISTORICAL INTRODUCTION}

From the beginning of times, the earth, sky, weather have been of great concern.
As soon as agriculture, commerce, travelling on land and sea prevailed, men have
wished to predict the weather. Later on airborne machines need atmosphere
knowledge and weather predictions for best flying. Nowadays there is much money
spent on weather predictions for sport activities. It is known how the knowledge
of  weather (temperature, wind, humidity, ..) is relevant, (even $fundamental$
!), e.g.  in sailing races or in Formula 1 and car rally races. Let it be
recalled the importance of knowing and predicting the wind (strength and
directions), pressure and temperature at high altitude for the (recent) no-stop
balloon round the world trip. The first to draw sea wind maps was
Halley\cite{bookmap}, an admirer of Breslau administration. That followed the
''classical'' isobaths and isoheights (these are geometrical measures !!!) for
sailors needing to go through channels.

I am very pleased to point out that Heinrich Wilhelm Brandes(1777-1834),
Professor of Mathematics and Physics at the University of Breslau was the
first\cite{bookmap} who had the idea of displaying weather data (temperature, air
pressure, a.s.o.) on geographical maps\footnote{It seems that H.W. Brandes left
Breslau to get his Ph.D. thesis in Heidelberg in 1826. Alas it seems that the
original drawings  are not available at this time. Where are they?}. Later  von
Humboldt (1769-1859)  had the idea to connect points in order to draw
isotherms\cite{bookmap}.    It is well known nowadays that various algorithms
will give various isotherms, starting from the same temperature data and
coordinate table.  In fact the maximum or minimum temperature as defined in
meteorology\cite{meteoTdef,Huschke} are far from the ones acceptable in physics
laboratories. Note that displayed isotherms  connect data points which values are
obtained at different times! No need to say that it seems essential to
concentrate on predicting the uncertainty in forecast models of weather and
climate as emphasized elsewhere\cite{Palmer}.

\section{ CLIMATE and WEATHER. The role of clouds}

Earth's climate is clearly determined by complex interactions between sun,
oceans, atmosphere, land and biosphere \cite{atmosphere,andrews}. The composition
of the atmosphere is particularly important because certain gases, including
water vapor, carbon dioxide, etc., absorb heat radiated from Earth's surface. As
the atmosphere warms up, it in turn radiates heat back to the surface that
increases the earth's ''mean surface temperature''.

Much attention has been paid recently \cite{physworld1,physworld2} to the
importance of the main components  of the atmosphere, in particular
clouds\cite{clouds}, in the water three forms --- vapor, liquid and solid, for
buffering the global temperature against reduced or increased solar
heating\cite{ou}. This leads to efforts to improve not only models of the earth's
climate but also predictions of climate change\cite{Hasselmann}, as understood
over long time intervals, in contrast to shorter time scales for weather
forecast. In fact, with respect to climatology the situation is very complicated
because one does not even know what the evolution equations are. Since controlled
experiments cannot be performed on the climate system, one relies on using ad hoc
models to identify cause-and-effect relationships. Nowadays there are several
climate models belonging to many different centers\cite{climatemodelgroups}.
Their web sites not only carry sometimes the model output used to make images but
also provide the source code. It seems relevant to point out here that the
stochastic resonance idea was proposed to describe climatology
evolution\cite{Benzi}.

It should be remembered that solutions of Navier-Stokes equations forcefully
depend  on the initial conditions, and steps of integrations. Therefore a great
precision on  the temperature, wind velocity, etc. cannot be expected and the
solution(s) are only looking like a mess after a few numerical
steps\cite{Pasini}. The Monte Carlo technique suggests to introduce successively
a set of initial conditions, perform the integration of the differential
equations and make an average thereafter\cite{Pasini}. It is hereby time to
mention Lorenz's\cite{Lorenz} work who simplified Navier-Stokes
equations searching for  some predictiability. 
However, predicting the outcome of such a	 set of equations with 
complex nonlinear interactions
taking place in an open system is a difficult task\cite{ramsey}.

The turbulent character in the atmospheric boundary layer (ABL) is one of its
most important features. Turbulence can be caused by a variety of processes, like
thermal convection, or mechanically generated by wind shear, or following
interactions influenced by the rotation of the Earth\cite{16,20}. This complexity
of physical processes and interactions between them create a variety of
atmospheric formations. In particular, in a cloudy ABL the radiative fluxes
produce local sources of heating or cooling within the mixed-layer and therefore
can greatly influence its turbulent structure and dynamics, especially in the
cloud base. Two practical cases, the marine ABL and the
continental ABL have been investigated for their scaling
properties\cite{nadia,DMWC96,MDWC97}

Yet, let it be emphasized that the first modern ideas of statistical physics
implemented on cloud studies through fractal $geometry$ are due to Lovejoy who
looked at the perimeter-area relationship of rain and cloud areas\cite{lovejoy1},
fractal dimension of their shape or ground projection. He discovered the
statistical self-similarity of cloud boundaries through area-perimeter analyses
of the geometry of satellites,fractal scaling of the cloud perimeter in the
horizontal plane. He found the fractal dimension $D_p  \simeq  4/3$ over a
spectrum of 4 orders of magnitude in size, for small fair weather cumuli ($\sim$
1021 km) up to huge stratus fields ($\sim$ 103 km).  Cloud size distributions
have also been studied from a scaling point of
view\cite{Cahalan82,116,neggers,rodts}. Rain has also received much
attenion\cite{pinho,andrade,miranda,tessier,73a,73b,73c}.

\section{ Modern Statistical Physics Approaches}

Due to the nonlinear physics laws governing the phenomena in the atmosphere, the
time series of the atmospheric quantities are usually
non-stationary\cite{Karner,Davis} as revealed by Fourier spectral analysis, -
whih is  usually the first technique to use. Recently, new techniques have been
developed that can systematically eliminate trends and cycles in the data and
thus reveal intrinsic dynamical properties such as correlations that are very
often masked by nonstationarities,\cite{r14,BA}. Whence many studies  reveal
long-range power-law correlations in geophysics time
series\cite{Davis,FraedrichBlender} in particular in
meteorology\cite{bunde1,bunde2,BTpaper,Tsonisetal1,Tsonisetal2,TalknerWeber,buda,40a}.
Multi-affine properties
\cite{MDWC97,wavelets,DM94,Davis2,genpol,66b,klono,kita,turbwavelet,mfatmoturb}
can also be identified, using singular spectrum or/and wavelets.

There are different levels of essential interest for sorting out correlations
from data, in order to increase the confidence in
predictability\cite{MalamudTurcotte}. There are investigations based on long-,
medium-, and short-range horizons. The $i$-diagram variability ($iVD$) method
allows to sort out some short range correlations. The technique has been used on
a liquid water cloud content data set taken from the Atlantic Stratocumulus
Transition Experiment (ASTEX) 92 field program \cite{astexiVD}.  It has also been
shown that the random matrix approach can be applied to the empirical correlation
matrices obtained from the analysis of the basic atmospheric parameters that
characterize the state of atmosphere\cite{RMTStatAtmCorr}. The principal
component analysis technique is a standard technique\cite{PCA} in meteorology and
climate studies. The Fokker-Planck equation for describing the liquid water path
\cite{kijats}  is also of interest. See also some tentative search for power law
correlations in the Southern Oscillation Index fluctuations characterizing El
Ni$\tilde{n}$o\cite{prenino}. But there are many other works of
interest\cite{68b}.

\subsection{Ice in cirrus clouds}

In clouds, ice appears in a variety of forms, shapes, depending on the formation
mechanism and the atmospheric conditions\cite{20,wavelets,18,210}. The cloud
inner structure, content, temperature, life time, .. can be studied. In cirrus
clouds, at temperatures colder than about $-40^{\circ}$~C ice crystals form.
Because of the vertical extent, ca. from about 4 to 14~km and higher, and the
layered structure of such clouds one way of obtaining some information about
their properties is mainly by using ground-based remote sensing
instruments\cite{westwater,westwater78,rees,ceilometer}. Attention can be
focussed\cite{40a} on correlations in the fluctuations of radar signals obtained
at isodepths of $winter$ and $fall$ cirrus clouds giving (i) the backscattering
cross-section, (ii) the Doppler velocity and (iii) the Doppler spectral width of
the ice crystals. They correspond to the physical coefficients used in Navier
Stokes equations to describe flows, i.e. bulk modulus, viscosity, and thermal
conductivity. It was found that power-law time correlations exist with a
crossover between regimes at about 3~to 5~min, but also $1/f$ behavior,
characterizing the top and the bottom layers and the bulk of the clouds. The
underlying mechanisms for such correlations likely originate in ice nucleation
and crystal growth processes.

\subsection{Stratus clouds}

In stratus clouds,  long-range power-law correlations
\cite{BTpaper,buda} and multi-affine properties\cite{DMWC96,MDWC97,kita}   have
reported  for the liquid water fluctuations, beside the spectral
density\cite{Gerber}. Interestingly, stratus cloud data retrieved from the
radiance, recorded as brightness
temperature,\footnote{http://www.phys.unm.edu/~duric/phy423/l1/node3.html} at the
Southern Great Plains central facility and operated in the vertically pointing
mode\cite{errorref}   indicated a Fourier spectrum, $S(f) ~\sim ~
f^{-\beta}$,  $\beta$ exponent equal to $1.56\pm 0.03$ pointing to a
nonstationary time series. The detrended fluctuation analysis (DFA) 
 method applied on the stratus
cloud brightness microwave recording\cite{BTpaper,JAM} indicates the existence of
long-range power-law correlations  over a two hour time.

Contrasts in behaviors, depending on seasons can be pointed out. The DFA analysis
of liquid water path data measured in April 1998  gives a scaling   exponent
$\alpha = 0.34 \pm 0.01$ holding from 3 to 60 minutes. This scaling range is
shorter than the 150~min scaling range\cite{BTpaper} for a stratus cloud in
January 1998 at the same site. For longer correlation times a crossover to
$\alpha=0.50 \pm 0.01$ is seen up to about 2 h, after which the statistics of the
DFA function is not reliable.

However a change in regime from Gaussian to non-Gaussian fluctuation regimes has
been clearly defined for the cloud structure changes using a finite size (time)
interval window. It has been shown that the DFA exponent turns from a low value
(about 0.3) to 0.5 before the cloud breaks. This indicates that the stability of
the cloud, represented by antipersistent fluctuations is (for some unknown reason
at this level)  turning into a system for which the fluctuations are similar to a
pure random walk. The same type of finding was observed for the so called Liquid
Water Path\footnote{The liquid water path (LWP) is the amount of liquid water in
a vertical column of the atmosphere; it is measured in cm$^{-3}$; ... sometimes in cm
!!!}.

The value of $\alpha \approx 0.3$ can be interpreted as the $H_1$ parameter of
the multifractal analysis of liquid water content\cite{DMWC96,MDWC97,DM94} and of
liquid water path \cite{kita}. Whence, the appearance of broken clouds and clear
sky following a period of thick stratus can be interpreted as a non equilibrium
transition or a sort of fracture process in more conventional physics. The
existence of a crossover suggests two types of correlated events as in classical
fracture processes: nucleation and  growth of diluted droplets.  Such a marked
change in persistence implies that specific fluctuation correlation dynamics
should be usefully inserted as ingredients in {\it ad hoc} models.

\subsection{Cloud base height}

The variations in the local $\alpha$-exponent (''multi-affinity'') suggest that
the nature of the correlations change with time, so called intermittency
phenomena. The evolution of the time series can be decomposed into successive
persistent and anti-persistent sequences. It should be noted that the
intermittency of a signal is related to existence of extreme events, thus a
distribution of events away from a Gaussian distribution, in the evolution of the
process that has generated the data. If the tails of the distribution function
follow a power law, then the scaling exponent defines the critical order value
after which the statistical moments of the signal diverge. Therefore it is of
interest to probe the distribution of the fluctuations of a time dependent signal
$y(t)$ prior investigating its intermittency.  Much work has been devoted to the
cloud base height\cite{genpol,66b,klono}, under various ABL conditions, and the
LWP\cite{kita,kijats}. Neither the distribution of the fluctuations of liquid
water path signals nor those of the cloud base height appear to be Gaussian. The
tails of the distribution follow a power law pointing to ''large events'' also
occurring in the meteorological (space and time) framework.  This may suggest
routes for other models.

\subsection{Sea Surface Temperature}

Other time series analysis have been
investigated searching for power law exponents, like in 
atmospheric\cite{Pelletier}  or sea surface
temperature (SST) fluctuations\cite{Monetti}. These are of importance for weighing
their impacts on regional climate, whence finally to greatly increase
predictability of precipitation during all seasons.
Currently, climate patterns derived from global 
 SST are used to forecast precipitation.  

Recently we have attempted to observe whether the fluctuations in the Southern
Oscillation index ($SOI$) characterizing El Ni$\tilde{n}$o were also prone to a
power law analysis. For the  $SOI$ monthly averaged data time interval 1866-2000,
the tails of the cumulative distribution of the fluctuations of $SOI$ signal it
is found that large fluctuations are more likely to occur than the Gaussian
distribution would predict. An antipersistent type of correlations exist for a
time interval ranging from about 4 months to about 6 years. This leads to favor
specific physical models for El Ni$\tilde{n}$o description\cite{prenino}.

\section{Conclusions}

Modern statistical physics techniques for analyzing atmospheric time series
signals indicate scaling laws (exponents and ranges) for correlations. A few
examples have been given briefly here  above, mainly from contributed papers in
which the author has been involved.  Work by many other authors have not been
included for lack of space. This brief set of comments is only intended for
indicating how meteorology and climate problems can be tied to scaling laws and
inherent time series data analysis techniques.  Those ideas/theories have allowed
me to reduce the list of quoted references, though even like this I might have
been unfair. One example can be recalled in this conclusion to make the point:
the stratus clouds break when the molecule density fluctuations become Gaussian,
i.e. when the molecular motion becomes Brownian-like. This should lead to better
predictability on the cloud evolution and enormously extend the predictability
range in weather forecast along the lines of nonlinear dynamics\cite{6}.

\vskip 0.6cm {\bf Acknowledgments} \vskip 0.6cm

Part of this studies have  been supported through an Action Concert\'ee Program
of the University of Li$\grave e$ge (Convention 02/07-293). Comments by A.
P\c{e}kalski, N. Kitova, K. Ivanova and C. Collette are greatly appreciated.

 \end{document}